\documentclass[pra,preprint,a4paper]{revtex4}
\usepackage{graphicx}

\begin{document}
\title{Reduced Density Matrix Functional for Many-Electron Systems}
\author{S. Sharma$^{1,2,3}$}
\email{sangeeta.sharma@physik.fu-berlin.de}
\author{J. K. Dewhurst$^4$} 
\author{N. N. Lathiotakis$^{2,5}$}
\author{E. K. U. Gross$^{2,3}$}
\affiliation{1  Fritz Haber Institute of the Max Planck Society, Faradayweg 4-6, 
D-14195 Berlin, Germany.}
\affiliation{2 Institut f\"{u}r Theoretische Physik, Freie Universit\"at Berlin,
Arnimallee 14, D-14195 Berlin, Germany}
\affiliation{3 European Theoretical Spectroscopy Facility (ETSF)} 
\affiliation{4 School of Chemistry, The University of Edinburgh, 
Edinburgh EH9 3JJ.}
\affiliation{5 Theoretical and Physical Chemistry Institute,
The National Hellenic Research Foundation,
Vass. Constantinou 48, 11635 Athens, Greece }\date{\today}

\begin{abstract}
Reduced density matrix functional theory for the case of solids is presented and
a new exchange correlation functional based on a fractional power of 
the density matrix is introduced. We show that compared to other functionals, this  
produces more accurate results for finite systems.
Moreover, it captures the correct band gap behavior for conventional 
semiconductors as well as strongly correlated Mott insulators, where a gap is 
obtained in absence of any magnetic ordering.

\end{abstract}

\pacs{71.10.-w, 71.27.+a, 71.45.Gm, 71.20.Nr}
\maketitle


One of the most dramatic failures of the usual local density approximation (LDA)
or generalized gradient type approximations to the exchange-correlation (xc) 
functional of density functional theory (DFT) is the incorrect prediction of a
metallic ground state for the strongly correlated Mott insulators, of which
transition metal oxides (TMOs) may be considered as prototypical.
For some TMOs (NiO and MnO) spin polarized calculations do show a very small 
band gap (up to 95\% smaller than experiments) but only \emph{as the result of 
AFM ordering}, however all TMOs are found to be metallic in a spin unpolarized 
treatment. On the other hand, it is well known experimentally that these 
materials are insulating in nature even at elevated temperatures (much above the 
N\'eel temperature) \cite{mott}.
This indicates that the magnetic order is not the driving mechanism for the
gap and is just a co-occurring phenomenon. A real challenge for any kind of 
{\it ab-initio} theory then is the prediction of an insulating state for these 
strongly correlated materials in the absence of magnetic order.
Until now the main focus of reduced density matrix functional theory (RDMFT) has 
been on finite systems like atoms and 
molecules \cite{mueller,GU,bb0,gritsenko,herbert,cga,kollmar,leiva,neks,nekgap}
with various xc 
functionals \cite{GU,bb0,gritsenko,herbert,cga,kollmar,leiva,csanyi,pernal04,piris,nekjel}
which are essentially modifications of the original M\"{u}ller 
functional \cite{mueller}. In the present work we extend RDMFT to the case of 
solid-state systems and introduce a new functional which generates not only 
accurate gaps for conventional semiconductors, but demonstrates insulating 
behavior for Mott-type insulators in the non-magnetic phase.

\begin{figure}[ht]
\centerline{\includegraphics[width=0.6\columnwidth,angle=-90]{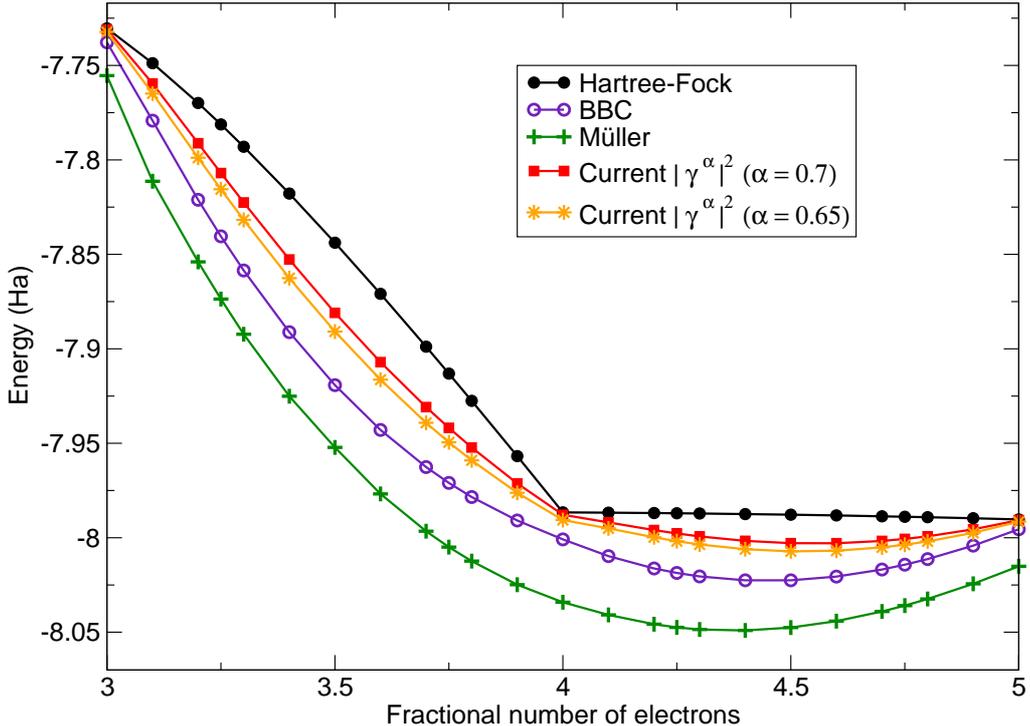}}
\caption{(Color online) Energy for LiH molecule as a function of fractional 
number of excess electrons. Results are obtained using various approximations 
to the xc functional.}
\label{tote}
\end{figure}

Formally, the one-body reduced density matrix $\gamma$ for a pure state of 
$N$ electrons is defined as (spin degrees of freedom are omitted for simplicity)

\begin{equation}
 \gamma({\bf r},{\bf r}')=N\int \Psi({\bf r},{\bf r}_2,\ldots,{\bf r}_N)
 \Psi^*({\bf r}',{\bf r}_2,\ldots,{\bf r}_N)\,d^3r_2\ldots d^3r_N.
\end{equation}
Diagonalization of this matrix produces a set of natural orbitals \cite{lodwin}
, $\phi_i$, and occupation numbers, $n_i$, leading to the spectral representation 

\begin{equation}
 \gamma({\bf r},{\bf r}')=\sum_i n_i \phi_i({\bf r})\phi_i^*({\bf r}')
\end{equation}
where the necessary and sufficient conditions for ensemble 
$N$-representability \cite{coleman} require $0\le n_i \le 1$ for all $i$, 
and $\sum_i n_i=N$. In terms of $\gamma$, the total ground state energy of the 
interacting system is \cite{gilbert} (atomic units are used throughout)


\begin{eqnarray}
 E_{\rm V}[\gamma]=\frac{-1}{2}\int\lim_{{\bf r}\rightarrow{\bf r}'}
 \nabla_{\bf r}^2\gamma({\bf r},{\bf r}')\,d^3r'+\int\rho({\bf r})
 V({\bf r})\,d^3r
+\frac{1}{2}\int
 \frac{\rho({\bf r})\rho({\bf r}')}{|{\bf r}-{\bf r}'|}\,d^3r\,d^3r'
 +E_{\rm xc}[\gamma],
\end{eqnarray}
where $\rho({\bf r})=\gamma({\bf r},{\bf r})$, $V$ is a given
external potential and $E_{\rm xc}$ we call the xc energy
functional. Minimizing the total energy with
$E_{\rm xc}[\gamma]=-\frac{1}{2}\int|\gamma({\bf r},{\bf r}')|^2/|{\bf r}-{\bf r}'|
\,d^3r\,d^3r'$ is equivalent to the Hartree-Fock (HF) method.
The HF functional satisfies the exact condition of the xc hole
integrating to minus one, however it does not satisfy the condition of 
convexity, which is required by the exact functional \cite{perdew}. M\"{u}ller 
proposed \cite{mueller} a simple alternative to the HF functional in which 
$|\gamma({\bf r},{\bf r}')|^2$ is replaced by 
$(\gamma^p({\bf r},{\bf r}'))^*\gamma^{1-p}({\bf r},{\bf r}')$,
where $\gamma^{\alpha}$ indicates the power used in the operator sense i.e.
\begin{equation}
\gamma^{\alpha}({\bf r},{\bf r}')=\sum_i n^{\alpha}_i \phi_i({\bf r})\phi_i^*({\bf r}') 
\end{equation}
M\"{u}ller's functional satisfies both conditions \cite{frank} for all $p$ with
$0<p<1$. All studies of 
this functional to date have set $p=1/2$ (in the rest of the paper we refer
to this as M\"{u}ller functional).
As is well known, however, this functional severely 
over-estimates electron correlation \cite{csanyi,gritsenko,herbert,nekjel} 
and we find that changing the value of $p$ away from $1/2$ only exacerbates this 
problem. Therefore in the present work we instead replace
$|\gamma({\bf r},{\bf r}')|^2$ by $|\gamma^{\alpha}({\bf r},{\bf r}')|^2$,
where $1/2\le\alpha\le 1$.
This simple functional form interpolates between the uncorrelated HF limit
($\alpha=1$) and the over-correlating M\"{u}ller functional
($\alpha=1/2$). Remarkably, we find that an intermediate value of $\alpha$ 
exists for which non-magnetic TMO's are insulators as well as accurate gaps
found for a diverse set of semiconductors and insulators.
The price to pay for this improved treatment of correlation is that this
functional form fails to satisfy the xc hole condition exactly \cite{xc-cond}.
However, all numerical tests show that it is at least convex and in fact, 
as shown in Fig. \ref{tote} for the case of finite systems (Fig. \ref{tote} is
a plot for a representative finite system, the LiH molecule), the variation of 
energy as a function of charge obtained using current functional 
(for $\alpha=0.65,0.7$) is much closer to the required straight-line 
behavior \cite{perdew} than the M\"{u}ller functional or the so called BBC
functionals \cite{gritsenko}. This type of linear behavior has also been 
recently demonstrated by Cohen {\it et al.} \cite{yang} for their density 
functional.

In  solids because of the
underlying lattice periodicity of the external potential,
the one-body reduced density matrix for crystals has the
symmetry $\gamma({\bf r}+{\bf T},{\bf r}'+{\bf T})=\gamma({\bf r},{\bf r}')$,
where ${\bf T}$ is a primitive translation vector, and thus
the natural orbitals are also Bloch states. 
Since there are no Kohn-Sham-like equations to solve, a direct minimization over
these natural orbitals and occupation numbers is required while maintaining the
ensemble $N$-representability conditions. In practical terms, the natural
orbitals are expanded in terms of a set of previously converged Kohn-Sham states,
and optimization of the natural orbitals is performed by varying the expansion
coefficients. We should stress that the use of Kohn-Sham 
states as a basis is merely a numerical convenience and results do 
not depend upon the starting point in any way.   
All calculations are performed using the state-of-the-art full-potential 
linearized augmented plane wave (FP-LAPW)
method \cite{Singh}, implemented within the EXCITING code \cite{exciting}.


For finite systems the chemical potential is defined as $\mu(N)= dE(N)/dN$
where $E(N)$ is the ground-state energy of $N$ electrons moving in a fixed
external potential. For the exact functional, $\mu(N)$ consists of horizontal 
lines with possible steps at integer values of $N$ \cite{perdew}. If $N_0$ is 
the electron number for which the total system is charge neutral, then the step
at $N_0$ given by $E_{\rm g}=\lim_{\eta\rightarrow 0^+} [\mu(N_0+\eta)-\mu(N_0-\eta)]$ 
represents the band gap or twice the chemical hardness. For periodic solids both 
$E$ and $N$ are infinite. Moreover, adding a finite charge $\eta$ to each unit-cell
while keeping the external potential fixed would lead to an infinitely charged
system which is not stable. The only accessible quantity is the energy per unit 
volume, $\tilde{E}$, as a function of excess charge $\eta$ per unit-cell, where 
$\tilde{E}_{{\rm v}+\delta v}(\eta)$ is the self-consistent total energy evaluated 
with an external potential (V) plus the Coulomb potential of a constant charge 
background ($\delta v$) which makes                     
the total system charge neutral. Clearly, the quantity $\tilde{E}$ and  
the chemical potential 
$\tilde{\mu}(\eta)= d\tilde{E}_{{\rm V}+\delta v}(\eta)/d\eta$ are
conceptually quite different from the corresponding quantities $E(\eta)$ and 
$\mu(\eta)$ of a finite system because for infinite systems the external potential
is not kept fixed as $\eta$ is changed. 
For the case of small excess charge $\eta$, the quantity  $\tilde{E}$  can be
written, to second order in $\eta$, as
\begin{equation}\label{Etilde1}
\tilde{E}_{{\rm V}+\delta v}(\eta)=\tilde{E}_{\rm V}(\eta) + \int \delta v({\bf r}) 
\tilde{\rho}_{N \pm \eta}({\bf r}) d^3r
\end{equation}       
The Coulomb potential from the constant charge ($\eta$) background is given by 
\begin{equation}\label{deltav}
\delta v({\bf r})= \eta \int \frac{1}{|{\bf r}-{\bf r}'|} d^3r'
\end{equation} 
and $\tilde{\rho}_{M}({\bf r})$, the self-consistent $M$-electron density  
evaluated with lattice potential plus neutralizing background potential, can
be expressed as
\begin{equation}\label{rhotilde} 
\tilde{\rho}_{N \pm \eta}({\bf r})=\tilde{\rho}_{N}({\bf r})\pm\eta n^{\pm}({\bf r})
\end{equation}
with $\int n^{\pm}({\bf r}) d^3r=1$.
Combining Eqs.(\ref{Etilde1}), (\ref{deltav}) and (\ref{rhotilde}) we arrive at
\begin{equation}\label{Etilde2}
\tilde{E}_{{\rm V}+\delta v}(\eta)= \tilde{E}_{\rm V}(\eta)\pm \eta \int 
\frac{\tilde{\rho}_{N}({\bf r})\pm\eta n^{\pm}({\bf r})}{|{\bf r}-{\bf r}'|} d^3r d^3r'. 
\end{equation}     
Derivative of Eq.(\ref{Etilde2}) with respect to $\eta$ yields
\begin{equation}\label{mulin}
\tilde{\mu}(\eta)=\mu(\eta=0^{-}) + \, \,  \, \, \, \, \left\{
\begin{array}{cr}
c_{\rm l} \eta & {\rm for} \, \, \eta < 0  \\
E_{\rm g}+c_{\rm r} \eta &  {\rm for} \, \, \eta > 0
\end{array}
\right.
\end{equation}
with $c_{\rm l}=2\int n^{-}({\bf r})/|{\bf r}-{\bf r}'| d^3r d^3r'$,    
$c_{\rm r}=2\int n^+({\bf r})/|{\bf r}-{\bf r}'| d^3r d^3r'$ and $\mu=d\tilde{E}_{\rm V}/d\eta$.
From Eq. (\ref{mulin}) we can see that introduction of the background charge has
no effect on the size of the band gap. Another aspect that is clear from this
equation is that unlike for the case of finite systems $\tilde{\mu}$ does not
consist of horizontal lines on either side of $\eta=0$.
Although Eq. (\ref{mulin}) is only valid for small values of $\eta$, 
in practice we find a linear behavior for $\tilde{\mu}(\eta)$ even for larger
values of $\eta$, a fact which can be attributed to the metallic nature that the
system acquires on addition or removal of charge.

\begin{figure}[ht]
\centerline{\includegraphics[width=0.8\columnwidth,angle=-0]{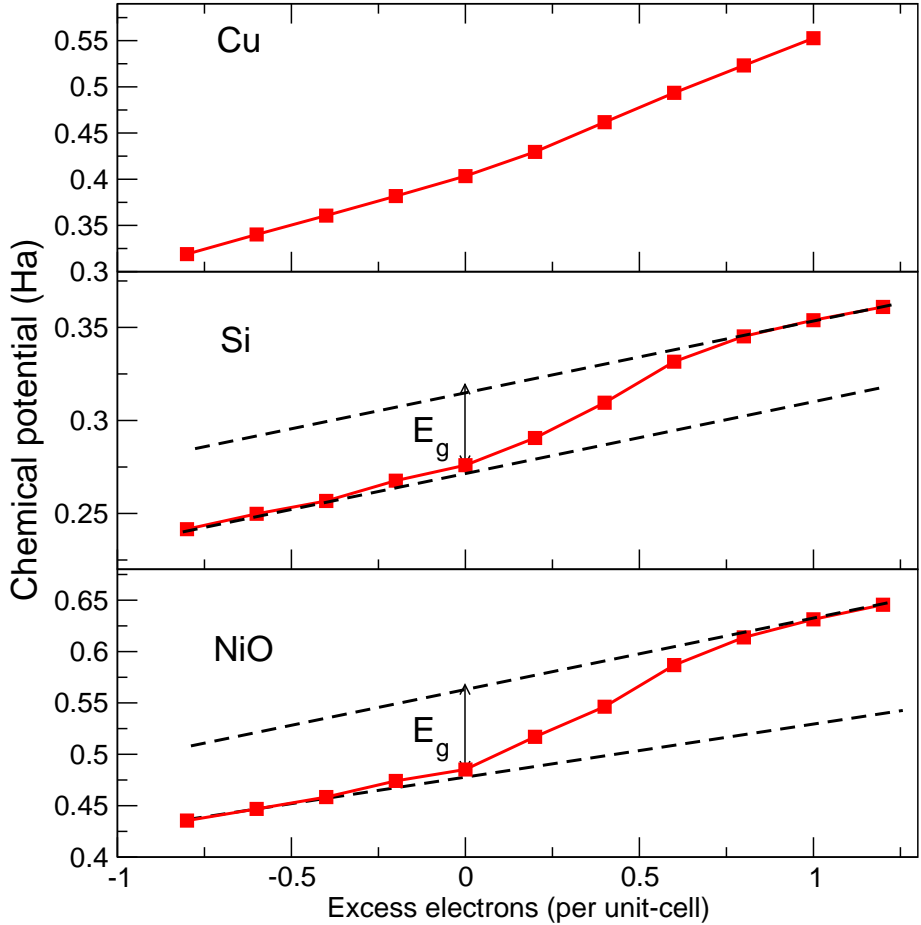}}
\caption{(Color online) Chemical potential (in Ha) versus the excess charge per 
unit-cell for (a) Cu, (b) Si and (2) NiO. The results are obtained using 
the current functional with $\alpha = 0.65$.}
\label{mu}
\end{figure}

In Fig. \ref{mu} we show the plot of the chemical potential versus excess electronic
charge for a prototype metal (Cu), semi-conductor (Si) and Mott insulator (NiO)
obtained using the current functional with this value of $\alpha = 0.65$.
For Cu one observes a nearly linear behavior with a small negative curvature. 
In contrast, Si and NiO show
a qualitatively different behavior in which the curvature of 
the chemical potential changes sign. We interpret this as the 
appearance of a smoothed discontinuity with a linear behavior to the left and 
to the right. This is strikingly different from that
of Cu, which remains metallic.
This smoothing of discontinuities, a consequence of approximating the
xc functional, has already been noted in work on finite systems \cite{nekgap}. 
Nevertheless, owing to the near linearity of the chemical potential on either 
side of zero excess charge (see Eq. (\ref{mulin})), the fundamental gap may 
be estimated \cite{nekgap} by construction of two tangents as shown in
Fig. \ref{mu}. This allows a rigorous test of the functional by comparison with 
experiment of the fundamental gap of a wide class of materials; semiconductors, 
insulators, and strongly correlated systems. We emphasize that
for the  M\"{u}ller and the BBC functionals, without self-interaction correction,
all these systems turn out to be metallic.

\begin{figure}[ht]
\centerline{\includegraphics[width=0.6\columnwidth,angle=-90]{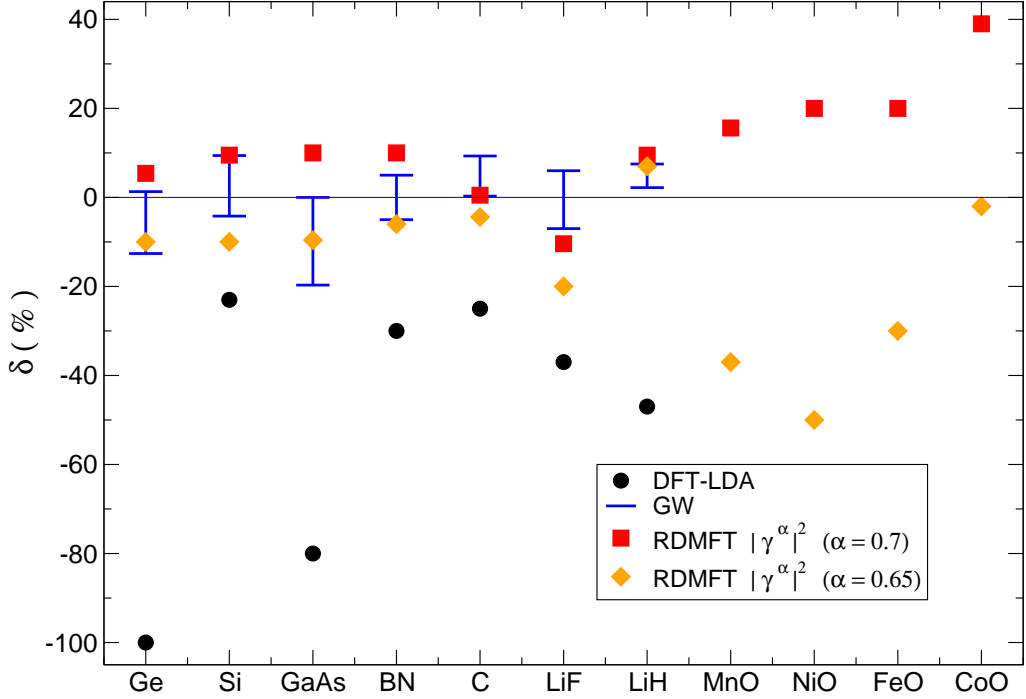}}
\caption{(Color online) Percentage deviation ($\delta$) of calculated band gap 
from experiment. DFT results (circles) are with the LDA. RDMFT values  
are obtained using the current functional with $\alpha = 0.65, 0.7$.
The $GW$ data (blue vertical lines) is
taken from Refs. [\onlinecite{gw}], a line is drawn 
between the smallest and the largest value.}
\label{eg}
\end{figure}

We first consider semiconductors and ionic insulators, for which results
are shown in Fig. \ref{eg}, again using the current functional with $\alpha = 0.65$.
The agreement between the calculated gap and experiments is good,
with an average deviation of the calculated gap 9.5\%, comparable to
the $GW$ method where, if we choose the best (worst) result from the literature 
for each material we find an average deviation of 3\% (9\%). It must be stressed 
that this good agreement of band gaps is found for a wide range of 
semiconductors/insulators, with gaps ranging from 1eV (Ge) to 14.2 eV (LiF) and 
the character of the material changing from predominantly covalent to ionic. 
We find that on increasing the value of $\alpha$ from 0.65 to 0.7 the agreement 
with experiment improves somewhat, with an average deviation of now 7.5\%. In 
most cases gaps are overestimated with $\alpha=0.7$ and underestimated with 
$\alpha=0.65$. Reducing the value of $\alpha$ below 0.65 has deleterious 
consequences; already for $\alpha=0.6$ several of the semiconductors are metallic. 
Therefore, there exists a small range of $\alpha$ (0.65-0.7) for which the 
calculated gaps agree well with experiment.

The TMOs have fundamental gaps ranging from 4.2 eV (NiO) 
to 2.4 eV (CoO), however the origin of the gap is profoundly different from
conventional semiconductors/insulators. Here the gap is opened by strong
Mott-Hubbard correlations and, towards the end of the 3$d$ series, charge 
transfer effects also contribute \cite{mott}. 
Gap formation by strong correlations is largely unaffected by the magnetic state, 
and so to focus exclusively on Mott-Hubbard physics we calculate these materials 
in the non-magnetic state. 
As can be seen in Fig. \ref{eg} the current functional (for both 
$\alpha = 0.65,0.7$) finds TMOs to be insulators. In this case the sensitivity 
of the gap to $\alpha$ is significantly greater,
however, as before, the best value of $\alpha$ is still in the range $0.65-0.7$.  

The traditional quasi particle technique used to treat TMOs, the $GW$ method, 
has only been applied to NiO and MnO in 
presence of AFM ordering. The reason for this lies in the fact that all DFT 
calculations to-date lead to metallic FeO and CoO, and for the $GW$ 
method to produce a gap starting from this metallic ground state would require a 
diverging  self-energy. 
In contrast, RDMFT is a fully {\it ab-initio} 
non-perturbative theory which with the current functional and a fixed value of 
$\alpha$, seems to capture not only the physics of 
conventional semiconductors/insulators, but also that of strongly correlated 
materials. What still remains is the method for choosing $\alpha$. It is clear 
from the present work that the optimal value of $\alpha$ lies in a small 
range ($0.65-0.70$). In order to refine $\alpha$ any further one could
do one of two things: make $\alpha$ a functional of the density matrix and 
determine this functional from exact properties of $E_{\rm xc}[\gamma]$ or fit $\alpha$
to best reproduce results for a wide range of materials.    
 
\begin{table}[t]
\begin{tabular}{c|cccc}
\hline\hline
Solid   & Expt. & DFT-LDA & M\"uller & current   \\ \hline
Diamond & 6.74  & 6.68    & 6.78     & 6.75      \\
Si      & 10.26 & 10.188  & 10.49    & 10.55     \\
BN      & 6.83  & 6.758   & 6.86     & 6.82      \\ 
	&	& 1.03    & 1.07     & 1         \\ \hline
\end{tabular}
\caption{\label{tab1} 
Equilibrium lattice parameter (in a.u.). The DFT values 
are obtained using the LDA. RDMFT results are obtained with the M\"uller and 
current functional with $\alpha =0.7$. In the last row are given the average 
percentage deviation of calculated results from experiments.}
\end{table}

It is clear that the current functional improves the value of the band gaps,
however it is also important that it performs well for properties already
adequately described by DFT within LDA. An example of such a property
is the equilibrium lattice constant. In Table \ref{tab1} are presented the
values for this for Si, cubic-BN and diamond. The results for the equilibrium 
lattice parameter obtained with RDMFT, using both M\"uller and current functional 
($\alpha = 0.7$), are as good as the values obtained using DFT. For the current
functional with $\alpha = 0.65$ the 
average percentage deviation of calculated results from experiments is 1.2\%.

In conclusion an extension of RDMFT to solids is presented with introduction 
of  a new functional. The values of the fundamental gap for various semiconductors
and ionic insulators are dramatically improved from DFT-LDA values and are in
as good agreement with experiment as $GW$ values. Furthermore, 
strong Mott-Hubbard correlation is captured in this approach, as calculations
of the non-magnetic TMOs show. The M\"uller functional appears
to be unsuitable for use with solids, as
it fails to produce a gap even for conventional semiconductors.

We have shown that RDMFT is a viable theory for the study of solids
where many body effects are important, and where DFT based theories 
have notably failed. This opens up many future possibilities, such as the 
study of High T$_{\rm C}$ superconductors in their under-doped Mott insulating
phase. It should also, hopefully, stimulate efforts to develop the theory
formally, including a temperature dependent extension
and a method to produce quasi particle spectra.

We acknowledge Deutsche Forschungsgemeinschaft and the Network of Excellence 
NANOQUANTA (NMP4-CT-2004-50019) for financial support.


\begin{thebibliography}{25}
\expandafter\ifx\csname natexlab\endcsname\relax\def\natexlab#1{#1}\fi
\expandafter\ifx\csname bibnamefont\endcsname\relax
  \def\bibnamefont#1{#1}\fi
\expandafter\ifx\csname bibfnamefont\endcsname\relax
  \def\bibfnamefont#1{#1}\fi
\expandafter\ifx\csname citenamefont\endcsname\relax
  \def\citenamefont#1{#1}\fi
\expandafter\ifx\csname url\endcsname\relax
  \def\url#1{\texttt{#1}}\fi
\expandafter\ifx\csname urlprefix\endcsname\relax\def\urlprefix{URL }\fi
\providecommand{\bibinfo}[2]{#2}
\providecommand{\eprint}[2][]{\url{#2}}

\bibitem[{mot()}]{mott}
\emph{\bibinfo{title}{{\rm O. Tjernberg et al., Phys. Rev. B {\bf 54}, 10245
  (1996), W. Jauch and M. Reehuis, Phys. Rev. B {\bf 70}, 195121 (2004)}}}.

\bibitem[{\citenamefont{M\"uller}(1984)}]{mueller}
\bibinfo{author}{\bibfnamefont{A.~M.~K.} \bibnamefont{M\"uller}},
  \bibinfo{journal}{Phys. Lett.} \textbf{\bibinfo{volume}{105A}},
  \bibinfo{pages}{446} (\bibinfo{year}{1984}).

\bibitem[{\citenamefont{Goedecker and Umrigar}(1998)}]{GU}
\bibinfo{author}{\bibfnamefont{S.}~\bibnamefont{Goedecker}} \bibnamefont{and}
  \bibinfo{author}{\bibfnamefont{C.~J.} \bibnamefont{Umrigar}},
  \bibinfo{journal}{Phys. Rev. Lett.} \textbf{\bibinfo{volume}{81}},
  \bibinfo{pages}{866} (\bibinfo{year}{1998}).

\bibitem[{\citenamefont{Buijse and Baerends}(2002)}]{bb0}
\bibinfo{author}{\bibfnamefont{M.~A.} \bibnamefont{Buijse}} \bibnamefont{and}
  \bibinfo{author}{\bibfnamefont{E.~J.} \bibnamefont{Baerends}},
  \bibinfo{journal}{Mol. Phys.} \textbf{\bibinfo{volume}{100}},
  \bibinfo{pages}{401} (\bibinfo{year}{2002}).

\bibitem[{\citenamefont{Gritsenko et~al.}(2005)\citenamefont{Gritsenko, Pernal,
  and Baerends}}]{gritsenko}
\bibinfo{author}{\bibfnamefont{O.}~\bibnamefont{Gritsenko}},
  \bibinfo{author}{\bibfnamefont{K.}~\bibnamefont{Pernal}}, \bibnamefont{and}
  \bibinfo{author}{\bibfnamefont{E.~J.} \bibnamefont{Baerends}},
  \bibinfo{journal}{J. Chem. Phys.} \textbf{\bibinfo{volume}{122}},
  \bibinfo{pages}{204102} (\bibinfo{year}{2005}).

\bibitem[{\citenamefont{Herbert and Harriman}(2003)}]{herbert}
\bibinfo{author}{\bibfnamefont{J.~M.} \bibnamefont{Herbert}} \bibnamefont{and}
  \bibinfo{author}{\bibfnamefont{J.~E.} \bibnamefont{Harriman}},
  \bibinfo{journal}{Chem. Phys. Lett.} \textbf{\bibinfo{volume}{382}},
  \bibinfo{pages}{142} (\bibinfo{year}{2003}).

\bibitem[{\citenamefont{Cs\'anyi et~al.}(2002)\citenamefont{Cs\'anyi,
  Goedecker, and Arias}}]{cga}
\bibinfo{author}{\bibfnamefont{G.}~\bibnamefont{Cs\'anyi}},
  \bibinfo{author}{\bibfnamefont{S.}~\bibnamefont{Goedecker}},
  \bibnamefont{and} \bibinfo{author}{\bibfnamefont{T.~A.} \bibnamefont{Arias}},
  \bibinfo{journal}{Phys. Rev. A} \textbf{\bibinfo{volume}{65}},
  \bibinfo{pages}{032510} (\bibinfo{year}{2002}).

\bibitem[{\citenamefont{Kollmar}(2004)}]{kollmar}
\bibinfo{author}{\bibfnamefont{C.}~\bibnamefont{Kollmar}}, \bibinfo{journal}{J.
  Chem. Phys.} \textbf{\bibinfo{volume}{121}}, \bibinfo{pages}{11581}
  (\bibinfo{year}{2004}).

\bibitem[{\citenamefont{Leiva and Piris}(2005)}]{leiva}
\bibinfo{author}{\bibfnamefont{P.}~\bibnamefont{Leiva}} \bibnamefont{and}
  \bibinfo{author}{\bibfnamefont{M.}~\bibnamefont{Piris}}, \bibinfo{journal}{J.
  Chem. Phys.} \textbf{\bibinfo{volume}{123}}, \bibinfo{pages}{214102}
  (\bibinfo{year}{2005}).

\bibitem[{\citenamefont{Lathiotakis et~al.}(2005)\citenamefont{Lathiotakis,
  Helbig, and Gross}}]{neks}
\bibinfo{author}{\bibfnamefont{N.~N.} \bibnamefont{Lathiotakis}},
  \bibinfo{author}{\bibfnamefont{N.}~\bibnamefont{Helbig}}, \bibnamefont{and}
  \bibinfo{author}{\bibfnamefont{E.~K.~U.} \bibnamefont{Gross}},
  \bibinfo{journal}{Phys. Rev. A} \textbf{\bibinfo{volume}{72}},
  \bibinfo{pages}{030501} (\bibinfo{year}{2005}).

\bibitem[{\citenamefont{Helbig et~al.}(2007)\citenamefont{Helbig, Lathiotakis,
  Albrecht, and Gross}}]{nekgap}
\bibinfo{author}{\bibfnamefont{N.}~\bibnamefont{Helbig}},
  \bibinfo{author}{\bibfnamefont{N.~N.} \bibnamefont{Lathiotakis}},
  \bibinfo{author}{\bibfnamefont{M.}~\bibnamefont{Albrecht}}, \bibnamefont{and}
  \bibinfo{author}{\bibfnamefont{E.~K.~U.} \bibnamefont{Gross}},
  \bibinfo{journal}{Europhys. Lett.} \textbf{\bibinfo{volume}{77}},
  \bibinfo{pages}{67003} (\bibinfo{year}{2007}).

\bibitem[{\citenamefont{Cs\'anyi and Arias}(2000)}]{csanyi}
\bibinfo{author}{\bibfnamefont{G.}~\bibnamefont{Cs\'anyi}} \bibnamefont{and}
  \bibinfo{author}{\bibfnamefont{T.~A.} \bibnamefont{Arias}},
  \bibinfo{journal}{Phys. Rev. B} \textbf{\bibinfo{volume}{61}},
  \bibinfo{pages}{7348} (\bibinfo{year}{2000}).

\bibitem[{\citenamefont{Pernal and Cioslowski}(2004)}]{pernal04}
\bibinfo{author}{\bibfnamefont{K.}~\bibnamefont{Pernal}} \bibnamefont{and}
  \bibinfo{author}{\bibfnamefont{J.}~\bibnamefont{Cioslowski}},
  \bibinfo{journal}{J. Chem. Phys.} \textbf{\bibinfo{volume}{120}},
  \bibinfo{pages}{5987} (\bibinfo{year}{2004}).

\bibitem[{\citenamefont{Piris}(2006)}]{piris}
\bibinfo{author}{\bibfnamefont{M.}~\bibnamefont{Piris}}, \bibinfo{journal}{Int.
  J. Quant. Chem} \textbf{\bibinfo{volume}{106}}, \bibinfo{pages}{1093}
  (\bibinfo{year}{2006}).

\bibitem[{\citenamefont{Lathiotakis et~al.}(2007)\citenamefont{Lathiotakis,
  Helbig, and Gross}}]{nekjel}
\bibinfo{author}{\bibfnamefont{N.~N.} \bibnamefont{Lathiotakis}},
  \bibinfo{author}{\bibfnamefont{N.}~\bibnamefont{Helbig}}, \bibnamefont{and}
  \bibinfo{author}{\bibfnamefont{E.~K.~U.} \bibnamefont{Gross}},
  \bibinfo{journal}{Phys. Rev. B} \textbf{\bibinfo{volume}{75}},
  \bibinfo{pages}{195120} (\bibinfo{year}{2007}).

\bibitem[{\citenamefont{L\"odwin}(1955)}]{lodwin}
\bibinfo{author}{\bibfnamefont{P.~O.} \bibnamefont{L\"odwin}},
  \bibinfo{journal}{Phys. Rev.} \textbf{\bibinfo{volume}{97}},
  \bibinfo{pages}{1474} (\bibinfo{year}{1955}).

\bibitem[{\citenamefont{Coleman}()}]{coleman}
\bibinfo{author}{\bibfnamefont{A.}~\bibnamefont{Coleman}},
  \emph{\bibinfo{title}{\rm{Rev. {M}od. {P}hys.} {\bf 35}, 668 (1963). {{T}his
  proof can be easily extended to the case of fractional number of electrons:
  Sharma et. al (preprint)}}}.

\bibitem[{\citenamefont{Gilbert}(1975)}]{gilbert}
\bibinfo{author}{\bibfnamefont{T.~L.} \bibnamefont{Gilbert}},
  \bibinfo{journal}{Phys. Rev. B} \textbf{\bibinfo{volume}{12}},
  \bibinfo{pages}{2111} (\bibinfo{year}{1975}).

\bibitem[{\citenamefont{{Perdew {\it et al.} }}(1982)}]{perdew}
\bibinfo{author}{\bibfnamefont{J.~P.} \bibnamefont{{Perdew {\it et al.} }}},
  \bibinfo{journal}{Phys. Rev. Lett.} \textbf{\bibinfo{volume}{49}},
  \bibinfo{pages}{1691} (\bibinfo{year}{1982}).

\bibitem[{\citenamefont{{Frank {\it et al.}}}(2007)}]{frank}
\bibinfo{author}{\bibfnamefont{R.~L.} \bibnamefont{{Frank {\it et al.}}}}
  (\bibinfo{year}{2007}), \urlprefix\url{http://arxiv.org/abs/0705.1587}.

\bibitem[{xc-()}]{xc-cond}
\emph{\bibinfo{title}{{\rm The violation of this condition is by a small
  amount. The integrated value of xc hole lies between 0.95-0.98 for most
  materials studied here.}}}

\bibitem[{\citenamefont{Cohen et~al.}(2007)\citenamefont{Cohen, Mori-Sanchez,
  and Yang}}]{yang}
\bibinfo{author}{\bibfnamefont{A.~J.} \bibnamefont{Cohen}},
  \bibinfo{author}{\bibfnamefont{P.}~\bibnamefont{Mori-Sanchez}},
  \bibnamefont{and} \bibinfo{author}{\bibfnamefont{W.}~\bibnamefont{Yang}}
  (\bibinfo{year}{2007}),
  \urlprefix\url{http://xxx.lanl.gov/find/cond-mat/0708.3175}.

\bibitem[{\citenamefont{Singh}(1994)}]{Singh}
\bibinfo{author}{\bibfnamefont{D.~J.} \bibnamefont{Singh}},
  \emph{\bibinfo{title}{\rm {Planewaves Pseudopotentials and the LAPW Method},
  {Kluwer Academic Publishers, Boston}}} (\bibinfo{year}{1994}).

\bibitem[{\citenamefont{Dewhurst et~al.}(2004)\citenamefont{Dewhurst, Sharma,
  and Ambrosch-Draxl}}]{exciting}
\bibinfo{author}{\bibfnamefont{J.~K.} \bibnamefont{Dewhurst}},
  \bibinfo{author}{\bibfnamefont{S.}~\bibnamefont{Sharma}}, \bibnamefont{and}
  \bibinfo{author}{\bibfnamefont{C.}~\bibnamefont{Ambrosch-Draxl}}
  (\bibinfo{year}{2004}), \urlprefix\url{http://exciting.sourceforge.net}.

\bibitem[{gw()}]{gw}
\emph{\bibinfo{title}{{\rm S. Massidda et al., Phys. Rev. B {\bf 55}, 13494
  (1997), M. Rohlfing et al. ,Phys. Rev. B {\bf 48}, 17791 (1993), Eric L.
  Shirley, Phys. Rev. B, {\bf 58}, 9579 (1998), M. Shishkin and G. Kresse,
  Phys. Rev. B, {\bf 75}, 235102 (2007), S. Lebegue et al., Euro. Phys. Lett.
  {\bf 63}, 562 (2003)}}}.

\end{thebibliography}

\end{document}